\documentclass[12pt]{article}
\usepackage{epsfig,psfrag}
\usepackage{textgreek}
\usepackage{dcolumn}
\usepackage{bm}
\usepackage{graphicx}
\usepackage{times}
\usepackage{amsmath}
\usepackage{color}
\usepackage{epstopdf}
\usepackage[utf8]{inputenc}
\usepackage[T1]{fontenc}
\usepackage{textcomp}
\usepackage{lmodern}
\usepackage[numbers,square,sort&compress]{natbib}
\usepackage{comment}

\setcitestyle{numbers,square}
\title{Room temperature surface emission on large-area photonic crystal quantum cascade lasers}

\author
{Yong Liang$^{1,\dagger}$, Zhixin Wang$^{1,\dagger,*}$, Johanna Wolf$^1$, Emilio Gini$^{2}$, 
\\
Mattias Beck$^1$, Bo Meng$^1$, J\'er\^ome Faist$^1$, and Giacomo Scalari$^1$
\\
\normalsize{$^1$: Institute for Quantum Electronics, ETH-Z\"urich, CH-8093 Z\"urich, Switzerland}
\\
\normalsize{$^2$: FIRST laboratory ETH-Z\"urich, Z\"urich 8093, Switzerland}
\\
\normalsize{$^{\dagger}$ Yong Liang and Zhixin Wang contributed equally to this work.}
\\
\normalsize{$^*$ Corresponding author: zhixwang@phys.ethz.ch}
}

\date{}

\begin{document} 

\baselineskip24pt

\maketitle 

\begin{abstract}
We design and fabricate large-area (1.1 mm $\times$ 1.1 mm) photonic crystal quantum cascade lasers, enabling single-mode (wavelength $\sim$ 8.5 \textmu m) surface emission at room temperature, with a  maximum peak power up to 176 mW. The beam divergence is $< 1 ^{\circ}$ and without side-lobes. Moreover, by introducing asymmetry into the photonic crystal pillar shape, a single-lobed far-field is realized. The photonic band structure is measured with high spectral (0.72 cm$^{-1}$) and angular (0.1 $^{\circ}$) resolution by using the photonic crystal quantum cascade laser itself as a detector.
\end{abstract}

\maketitle

Quantum cascade lasers (QCLs) are semiconductor laser sources that operate in the both mid-infrared  \cite{faist1994quantum} and terahertz region \cite{williams2007terahertz}. Nowadays, QCLs are the sources of choice for many laser-based applications, e.g., trace gas spectroscopy \cite{jagerska2014dual}, process control \cite{lang2010situ}, and biological sensing \cite{reyes2014study}. Surface emitting lasers are advantageous for their beam shape and the ease of two-dimensional (2D) integration \cite{hirose2014watt}. Based on intersubband transitions, the selection rule of QCL determines the light to be transverse magnetic (TM) polarized. Therefore, conventional vertical cavity surface emission design is not suitable for QCLs. In order to realize a surface-emitting QCL, novel coupling structures have been implemented, for example, photonic crystal (PhC) \cite{colombelli2003quantum,xu2010surface,diao_continuous-wave_2013,yao201510}, second-order distributed feedback (DFB) grating \cite{hofstetter1999surface,jouy2015surface}, microdisk and ring resonators \cite{mujagic2008low,mahler2009vertically,razeghi2017recent}. Compared to these other approaches, PhC-QCLs offer the possibility of achieving single mode operation together with narrow diverging surface emission, thanks to the unique 2D in-plane coupling mechanism and the Bragg reflection \cite{wang2017analytical}. As expected, a narrow beam requires a large device size in both in-plane dimensions. However, room temperature, single mode, surface-emission from a large-area PhC-QCL remains challenging. 
Previously, our group has demonstrated room temperature, single mode PhC-QCLs by using deep-etched buried heterostructure \cite{peretti2016room}. However, the laser output was extracted from the edge facet. Limited by both the optical overlap factor with the active region, as well as fabrication imperfections, the lasing threshold was as high as 7.8 kA/cm$^2$. Here, by combining the buried heterostructure approach with optimized dry-etching in the fabrication process, we present a large-area ($400$ periods in each in-plane dimension) surface-emitting PhC-QCL.

\begin{figure}[htbp]
\centering
\includegraphics[width=\linewidth]{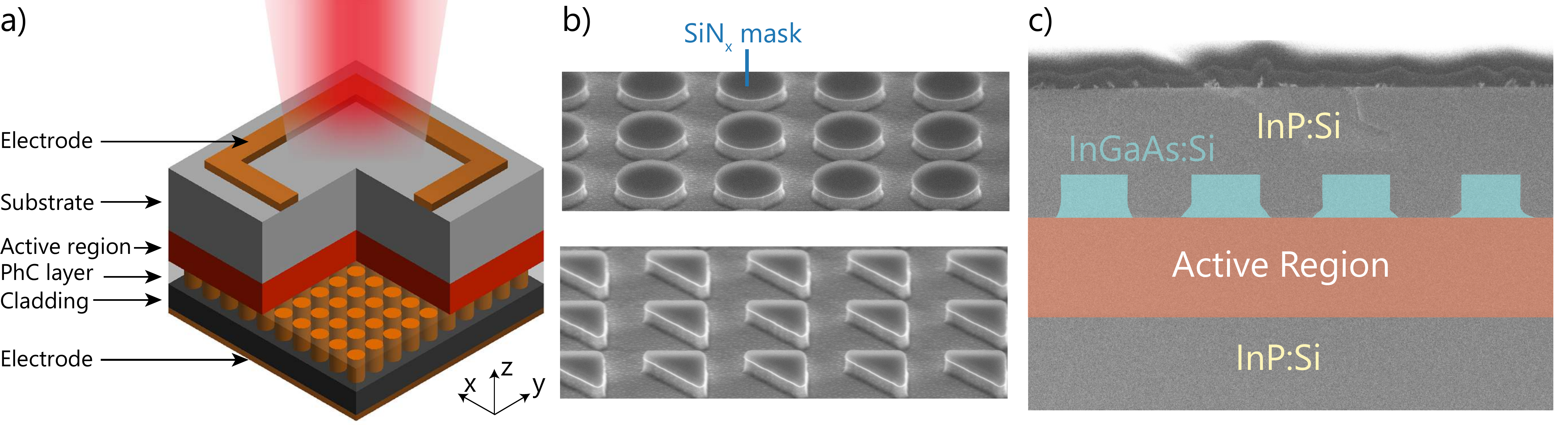}
\caption{(a) A schematic drawing of the surface-emitting PhC-QCL. (b) SEM images of the PhC-QCLs after ICP dry-etching. (c): SEM cross-sectional image of the PhC-QCL after InP cladding regrowth.}
\label{fig:1}
\end{figure}

The PhC layer of the laser is designed with an index contrast between 3.06 (InP) and 3.41 (InGaAs). The PhC is fabricated without etching the active region layer, in order to obtain a large optical overlap factor with the active region ($ \Gamma=\iiint_{AR} E_z^2 dV/\iiint_{All} E_z^2 dV \approx 56 \%$ when the filling factor is 0.40). According to our theoretical model, the 3D coupled wave theory \cite{wang2017analytical}, the $|\kappa L|$ is around 1.2 with such a PhC-QCL (400 periods $\times$ 400 periods). A schematic drawing of the designed PhC-QCL structure is shown in Fig. \ref{fig:1}(a). The active region (with a thickness of 2.05 \textmu m) is fabricated with lattice-matched InGaAs/AlInAs materials via molecular beam epitaxy (MBE) growth. The emission wavelength is approximately 8.5 \textmu m \cite{bismuto2010electrically}. A Si-doped InGaAs layer (800 nm thick) is then grown by metal-organic vapor phase epitaxy (MOVPE). After deep-ultraviolet lithography (220 nm wavelength), the layer is etched into PhC patterns by the inductively coupled plasma (ICP) dry-etching technique. We design square-lattice PhCs with two different pillar shapes: circle and right-angled isosceles triangle \cite{wang2017analytical,hirose2014watt}. The scanning electron microscopy (SEM) images of the them after the dry etching are shown in Fig. \ref{fig:1}(b). For the purpose of smooth and vertical sidewalls, SiN$_x$ hard mask is used for the ICP process, and is removed by HF wet etching later. The Si-doped-InP (InP:Si) cladding is then grown by MOVPE. The size of the laser cavity is defined by wet-etching the InP:Si cladding layer into square mesas with a dimension of around 1.1 mm $\times$ 1.1 mm (400 periods in both directions). Outside the mesa, a layer of SiN$_x$ is deposited, which has a high loss at $8.5$ \textmu m, and naturally creates an absorbing boundary in favor of the single-mode operation \cite{colombelli2003quantum}. The cross-sectional view of the device is shown in Fig. \ref{fig:1}(c). It is clear that the PhC pillars are completely buried by InP:Si without any air-hole voids and visible defects. After processing, the laser is epi-down mounted on an AlN submount, and the power is extracted from the substrate side.

\begin{figure}[htbp]
\centering
\includegraphics[width=0.75\linewidth]{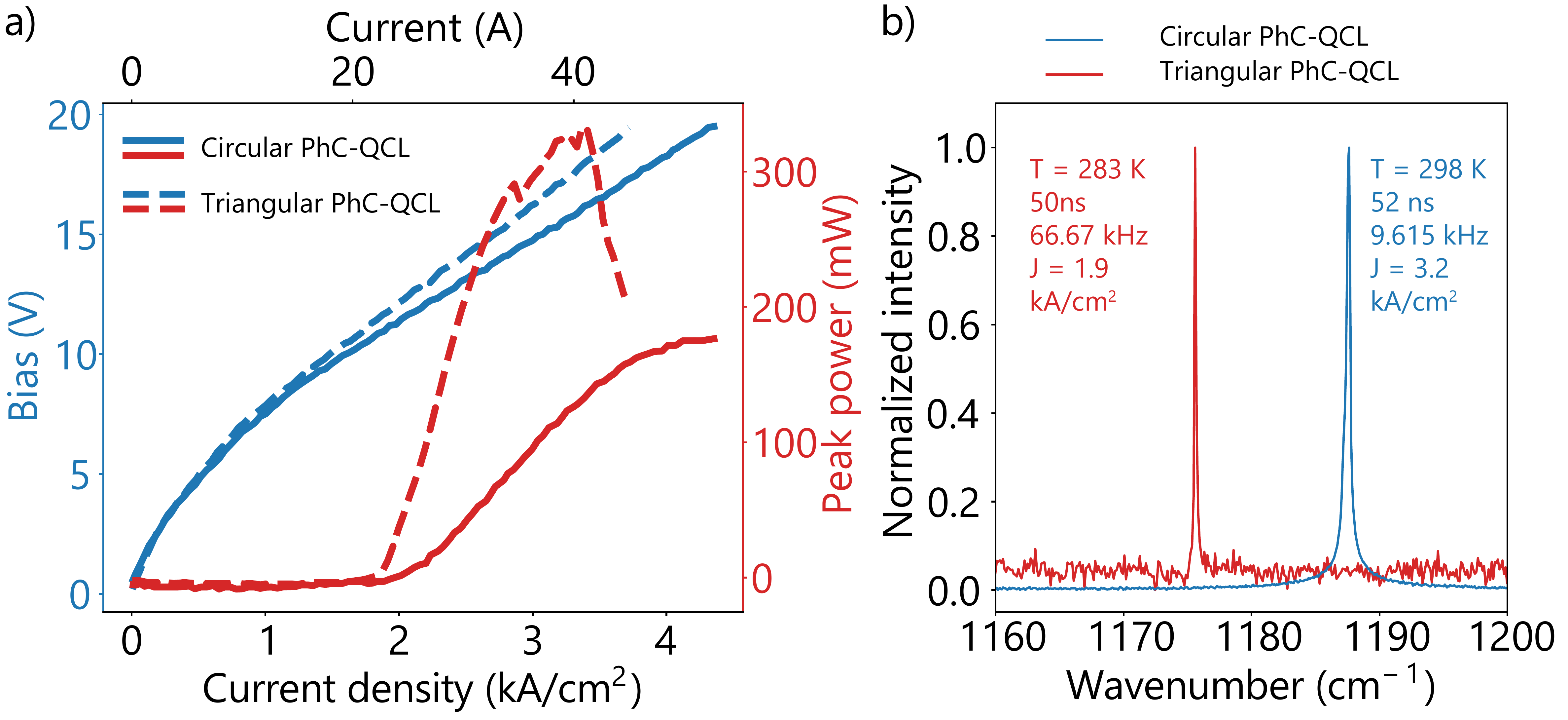}
\caption{(a) LIV characteristics of PhC-QCLs with circular shaped pillars and triangular shaped pillars. The measurement of the circular PhC-QCL is taken at room temperature (298 K) in pulsed operation (52 ns long pulse at 9.615 kHz). The measurement of the triangular PhC-QCL is taken under 284 K with 50 ns pulse at 66.67 kHz. The power is measured from the surface. The structural parameters are: circular PhC-QCL: period $= 2.680$ \textmu m, filling factor $= 0.40$; triangular PhC-QCL: period $= 2.710$ \textmu m, filling factor $= 0.28$. (b) Measured spectra of the same PhC-QCLs as (a). Driving conditions are shown in the inset.}
\label{fig:2}
\end{figure}

The light-current-voltage (LIV) characteristics (at pulsed operation, collected through the surface) of our lasers are presented in Fig. \ref{fig:2}(a), including the results of PhC-QCLs with circular shaped pillars (see the upper panel of Fig. \ref{fig:1} (b), measured at 298 K) and triangular shaped pillars (the lower panel of Fig. \ref{fig:1} (b), measured at 284 K). The threshold current density (J$_{th}$) is around 2 kA/cm$^2$ at room temperature. This low laser threshold is attributed to both the high overlap factor and the buried-heterostructure process. Figure \ref{fig:2}(b) shows the spectra of the corresponding lasers, measured by Fourier-transform infrared spectrometers (FTIRs) with the resolution of  0.075 cm$^{-1}$ (circular PhC-QCL) and 0.125 cm$^{-1}$ (triangular PhC-QCL), respectively. At room temperature, the circular shaped PhC-QCL is single-mode until the current density of 3.2 kA/cm$^2$ (the corresponding peak power is 110 mW). The single mode stability of the triangular shaped QCL is worse (single mode operation only retains near the threshold). This is partially because the asymmetrical structure is more sensitive to fabrication imperfections introduced by the dry etching. Besides, the filling factor (0.28) is smaller, leading to a worse index contrast. The maximum peak powers in Fig. \ref{fig:2}(a) are 176 mW (circular PhC-QCL) and 333 mW (triangular PhC-QCL). The slope efficiency dP/dI of them are 7.8 mW/A (circular PhC-QCL) and 17.5 mW/A (triangular PhC-QCL) .

The major reason of the low slope efficiency is the non-uniform electrical pumping of the active region. Compared to the device length (1.1 mm), the width of top electrode (100 \textmu m) and the thickness of the substrate (200 \textmu m) are both relatively small. Moreover, the doping of InP:Si substrate is only 1.5 $\times 10 ^ {16}$ cm$^{-3}$. A COMSOL simulation (see Supplement, Sec. \ref{sec:COMSOL}) indicates that under a bias of 20 V, the injected current density near the edge is around 1.64 kA/cm$^2$, whereas at the center it is only 0.46 kA/cm$^2$. This means almost 66\% of the active region (in the center) can be still below threshold, absorbing the power. Such non-uniform pumping also detriments the lateral mode control and the single mode stability. Besides, the dominating electric field component inside the QCL ($E_z$) is orthogonal to the electric field of the surface-emitted beam ($E_{x,y}$). Therefore, the surface coupling efficiency of TM modes is inherently weaker than TE modes \cite{yang2014three}. Besides, the edge emission power is not eliminated, despite the adoption of the absorbing boundaries (an edge emission power characteristic is shown in the Supplement, Sec. \ref{sec:edge}).

\begin{figure}[htbp]
\centering
\includegraphics[width=0.75\linewidth]{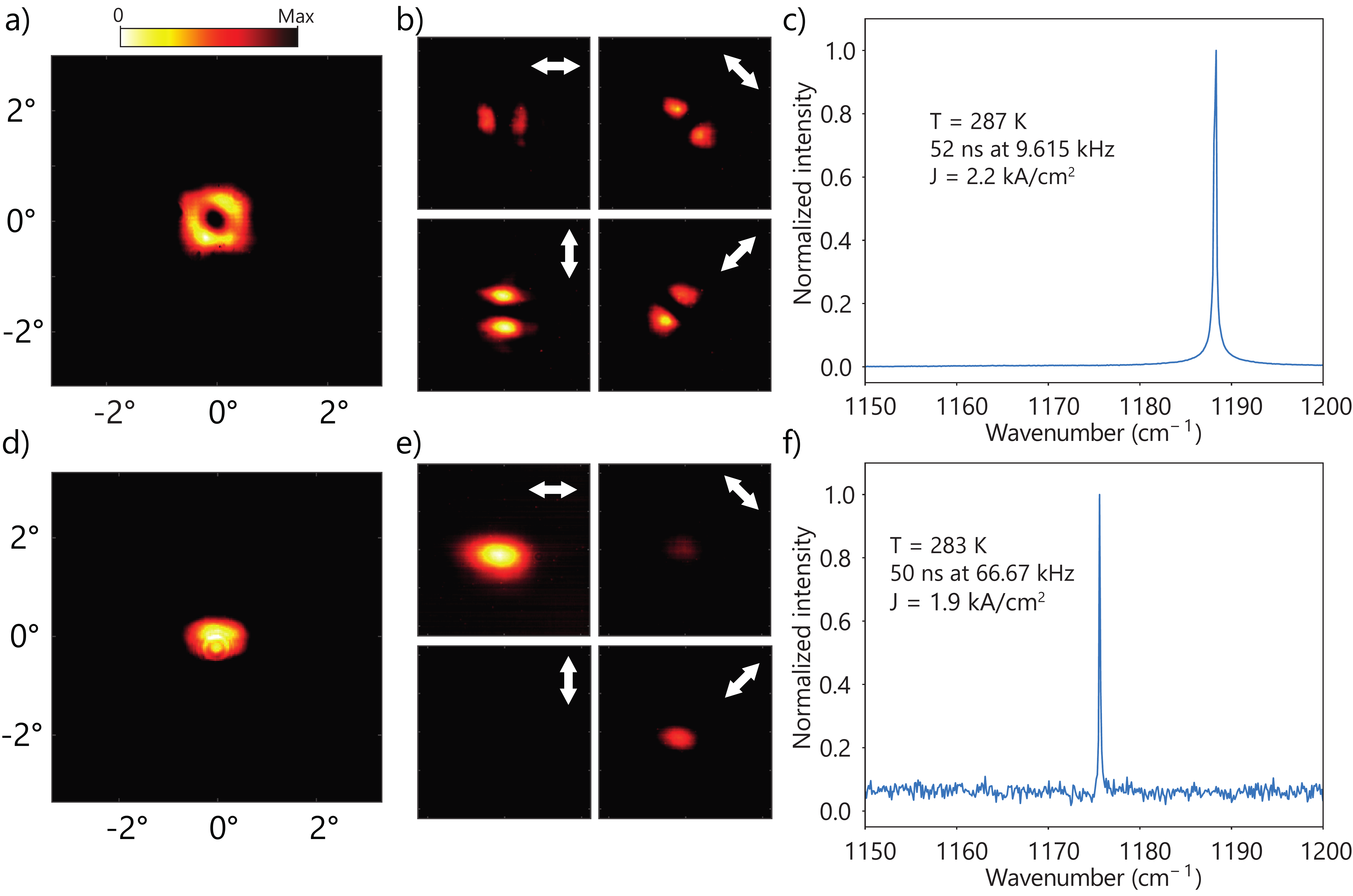}
\caption{Far-field patterns (a,d), polarization profiles in four typical directions (b,e), and the corresponding spectra (c,f) of the PhC-QCLs with circular pillars (a,b,c) and isosceles triangular pillars (d,e,f). The white arrows in (b,e) represent the direction of the polarization in each case. The structural parameters of the PhC-QCLs are: (a,b,c): period $= 2.680$ \textmu m, filling factor $= 0.40$; (d,e,f): period $= 2.710$ \textmu m, filling factor $= 0.28$ (both are the same lasers as shown in Fig. \ref{fig:2}).}
\label{fig:3}
\end{figure}

The surface-emitted far-field pattern of the circular-shaped PhC-QCL 
is shown in Fig. \ref{fig:3}. Figure \ref{fig:3}(b) illustrates the polarization profiles. Driving conditions and the corresponding spectrum are shown in Fig. \ref{fig:3}(c). The surface-emitted beam has an extremely small divergence angle ($< 1^{\circ}$) in all directions with no visible side lobes, owing to the large-area single-mode oscillation.

At the $\Gamma_2$ point of the square-lattice PhC-QCL, there are four band edge modes: two doubly degenerate dipole modes, a monopole mode and a quadrupole mode \cite{wang2017analytical}. For the PhC-QCL shown in Figs. \ref{fig:3}(a) and (b), 3D coupled wave theory indicates that the mode with the lowest cavity loss is the quadrupole mode which has a doughnut-shaped far-field, due to the destructive interference of the electric field emitted in the vertical direction \cite{wang2017analytical}. Besides, the model also predicts the polarization to be radial, as a typical feature of the TM mode. As we can see, the experimental results in Figs. \ref{fig:3}(a) and (b) agree well with the theoretical predictions. Experimentally, the beam can be observed without any lens by an infrared camera (aperture 10.88 mm $\times$ 8.16 mm) at a distance of 30 cm away from the PhC-QCL, keeping the same shape. 

In comparison, the surface far-field pattern of the PhC-QCL with isosceles-triangular pillars is shown in Fig. \ref{fig:3}(d). Figures \ref{fig:3}(e) and (f) are the corresponding polarization profiles, the spectrum and the driving conditions. 
The far-field is also remarkably narrow. Nevertheless, by breaking the in-plane symmetry of the internal field distribution, the destructive interference of the electric field is suppressed. Therefore, a single-lobed beam is observed. In this case, the beam is linearly polarized along one direction. 

\begin{figure}[htbp]
\centering
\includegraphics[width=\linewidth]{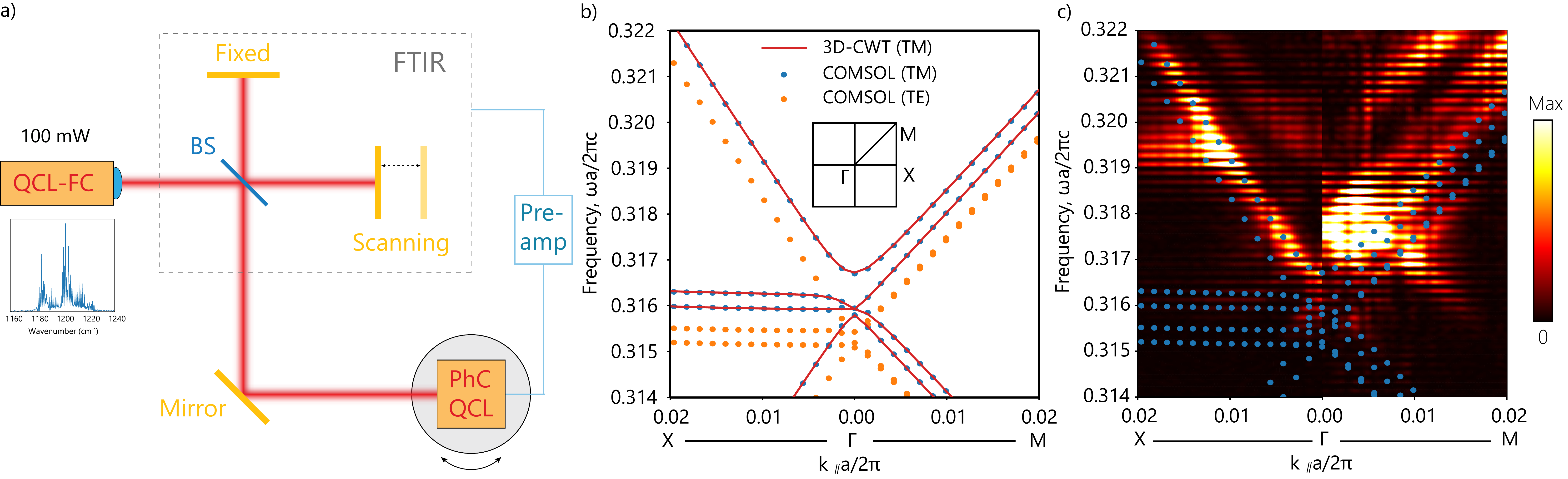}
\caption{(a) A schematic drawing of the band structure measurement setup. (b) Band structure of the PhC-QCL calculated by COMSOL Multiphysics and 3D coupled wave theory. The structural parameters are: period $= 2.680$ \textmu m, filling factor $= 0.40$ with circular-shaped pillars. $\Gamma$, X and M points of the reciprocal space are illustrated in the inset. (c) The combination of measured band structures along $\Gamma$ - X and $\Gamma$ - M directions, overlapped with COMSOL simulation data. The structural parameters are the same as (b).}
\label{fig:4}
\end{figure}

The photonic band structure is a crucial property of PhC structures. Previously, the most conventional way to measure the band on a PhC laser is by driving the laser below threshold and collecting the angular-resolved luminescence \cite{hirose2014watt}. We have implemented this approach on a PhC-QCL \cite{peretti2016room}, but the resolution is limited by the low radiation efficiency of intersubband transitions below threshold ($\tau^{-1}_{rad}/\tau^{-1}_{non-rad} \approx 10^{-5}$). To overcome this limitation, we establish a technique where the PhC-QCL is used as a photo-detector \cite{hofstetter2002quantum} and a band structure is mapped by photo-current measurement \cite{schartner2006band}.


A schematic drawing of our band measurement setup is described in Fig. \ref{fig:4}(a). A QCL frequency comb \cite{faist2016quantum} ($\sim$ 100 mW power, 46 cm$^{-1}$ bandwidth, continuous wave operation) is used as an illuminating laser source. The broadband spectrum of this laser source (as shown in the inset of Fig. \ref{fig:4}(a)) matches the lasing wavelength of the PhC-QCL. The collimated laser beam is directed into the FTIR. After traveling through the Michelson interferometer of the FTIR, the beam is incident on the surface of the PhC-QCL, generating a photo-current. The amplified photo-current is used as the input signal of the FTIR. To vary the angle between the PhC-QCL relative to the incident beam, the PhC-QCL is mounted on a rotation stage. The spectral range of such approach is the overlap between the gain of the QCL comb and the absorption bandwidth of the PhC-QCL detector.

The band structure is measured with the circular-pillar PhC-QCL (the same circular PhC-QCL characterized above). The bands of such a structure are calculated with our 3D coupled wave theory, and also simulated with the finite element method (COMSOL Multiphysics), as shown in Fig. \ref{fig:4}(b). The result of the 3D coupled wave theory agrees extremely well with the simulation. 

Since the PhC is designed into square lattices, both $\Gamma$-X and  $\Gamma$-M directions need to be measured to explore a whole band of the Brillouin zone. Experimentally, after the sweeping in $\Gamma$-X direction, the PhC-QCL is rotated in-plane by $45^{\circ}$, to scan along the $\Gamma$-M direction. Figure \ref{fig:4}(c) is the combination of both  $\Gamma$-X and  $\Gamma$-M  measurements (see Supplement Sec. \ref{sec:bandmeas} for individual results), overlapped by the COMSOL simulation. Near the $\Gamma$ point, the results in the two directions are not continuous. This is probably because the incident polarization is not the same. At the $\Gamma$ points of the two measurements, the in-plane wavevectors are both zero, but the polarization of incident electric field in the PhC plane is rotated by $45^{\circ}$ (see Supplementary Sec. \ref{sec:bandmeas} for a validation).

We attribute the shift between the theory and experimental measurements to our imperfect knowledge of the refractive indexes. Since the vertical symmetry of this structure is broken, the transverse electric (TE) modes also have $E_z$ components, i.e. there is a relatively strong coupling between the TE and TM modes. Therefore, the TE modes can also appear on the band. Besides, in Fig. \ref{fig:4}(b), only the first-order fundamental modes are taken into consideration as an ideal case. In comparison, there are additional modes shown in in Fig. \ref{fig:4}(c), which are likely the higher order waveguide modes. 

In this work,  large-area ($1.1$ mm $\times 1.1$ mm) surface-emitting PhC-QCLs are presented. The laser works at room temperature with a narrow far-field beam (< 1$^{\circ}$). In the band measurements, the angular step is 0.1$^{\circ}$ ($5 \times 10^{-4}$ in normalized wavevector, $k a/2\pi c$). The horizontal linewidth of the band is around $0.003$ , which has the same order of magnitude as the diffraction limit due to the size of the device ($0.001$) \cite{wang2016mode} and the divergence of the incident beam ($\sim 0.003$). The bands are also broadened because of the finite photon lifetime. For example, a 3 cm$^{-1}$ cavity loss leads to a frequency uncertainty $\Delta \omega = 0.0025$.  The resolution of the FTIR is 0.25 cm$^{-1}$ (7.5 GHz, $7 \times 10^{-5}$ in normalized frequency, $\omega a/2\pi c$). The spectral resolution of the band measurement is limited by the repetition frequency of our QCL comb (0.72 cm$^{-1}$, $2 \times 10^{-4}$ $\omega a/2\pi c$). All of these parameters thus ensure the highest resolution measurement of band structure on PhC-QCLs. In the end, it is our belief that there is  still a large room for improvement to optimize the performance of PhC-QCLs, and the large-area PhC-QCLs will eventually play a significant role in both industry and research applications.

\section*{Acknowledgement}

This work was financially supported by H2020 European Research Council Consolidator Grant CHIC 724344 and FP7 People: Marie-Curie Actions (FEL-27 14-2). The authors also thank Filippos Kapsalidis, Matthew Singleton, Mehran Shahmohammadi, Ilia Sergachev and Chao Peng for technical supports and fruitful discussions.

%

\newpage
\section*{Supplemental Documents}


\subsection{COMSOL simulation on the non-uniform electrical pumping}
\label{sec:COMSOL}

\begin{figure}[htbp]
\centering
\includegraphics[width=0.75\linewidth]{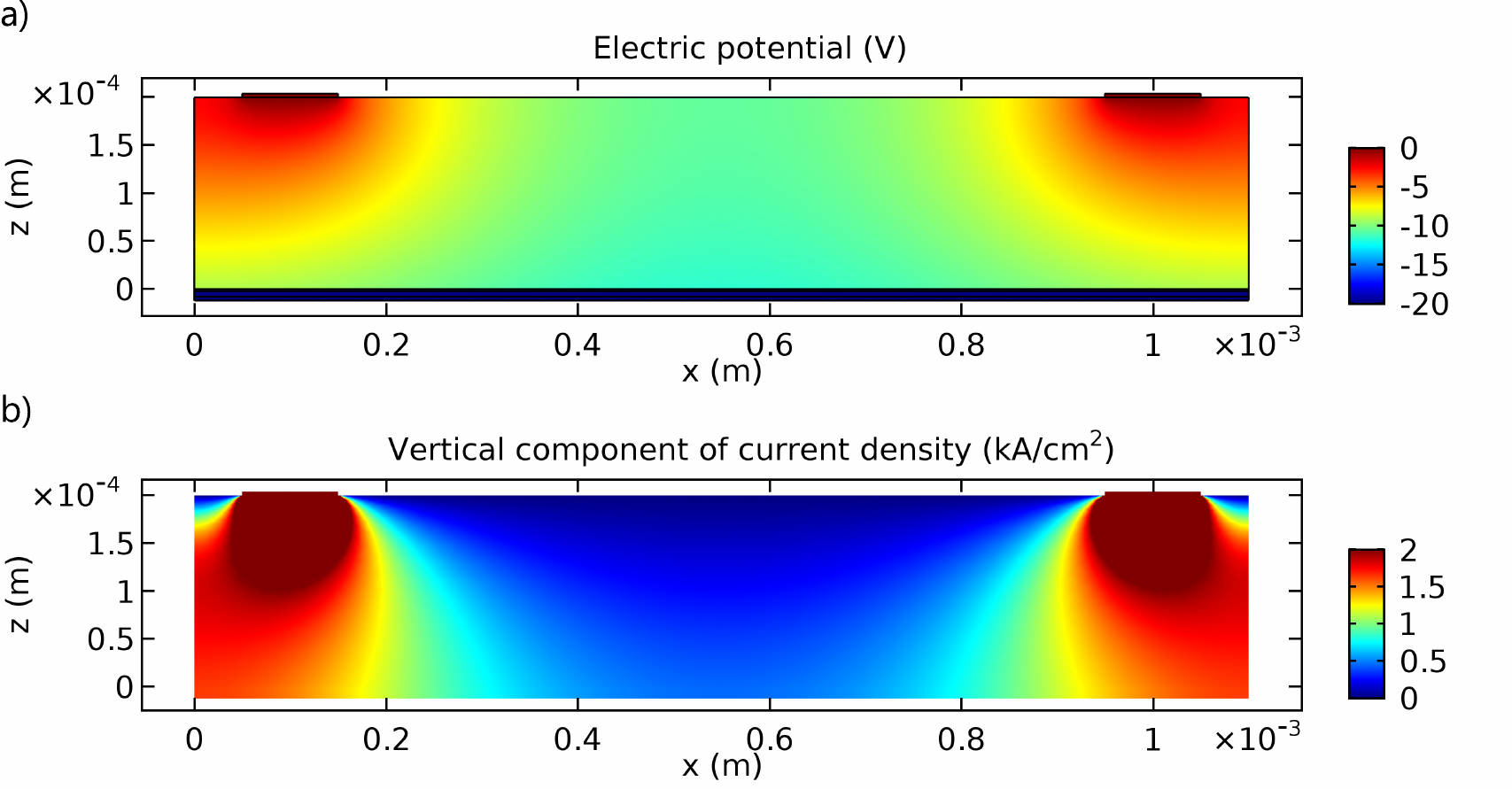}
\caption{The distribution of electric potential (a) and vertical component of current density $J_z$ (b) in the cross-sectional plane , simulated by COMSOL Multiphysics, with a bias of 20 V.}
\label{fig:COMSOL_2D}
\end{figure}

\begin{figure}[htbp]
\centering
\includegraphics[width=0.5\linewidth]{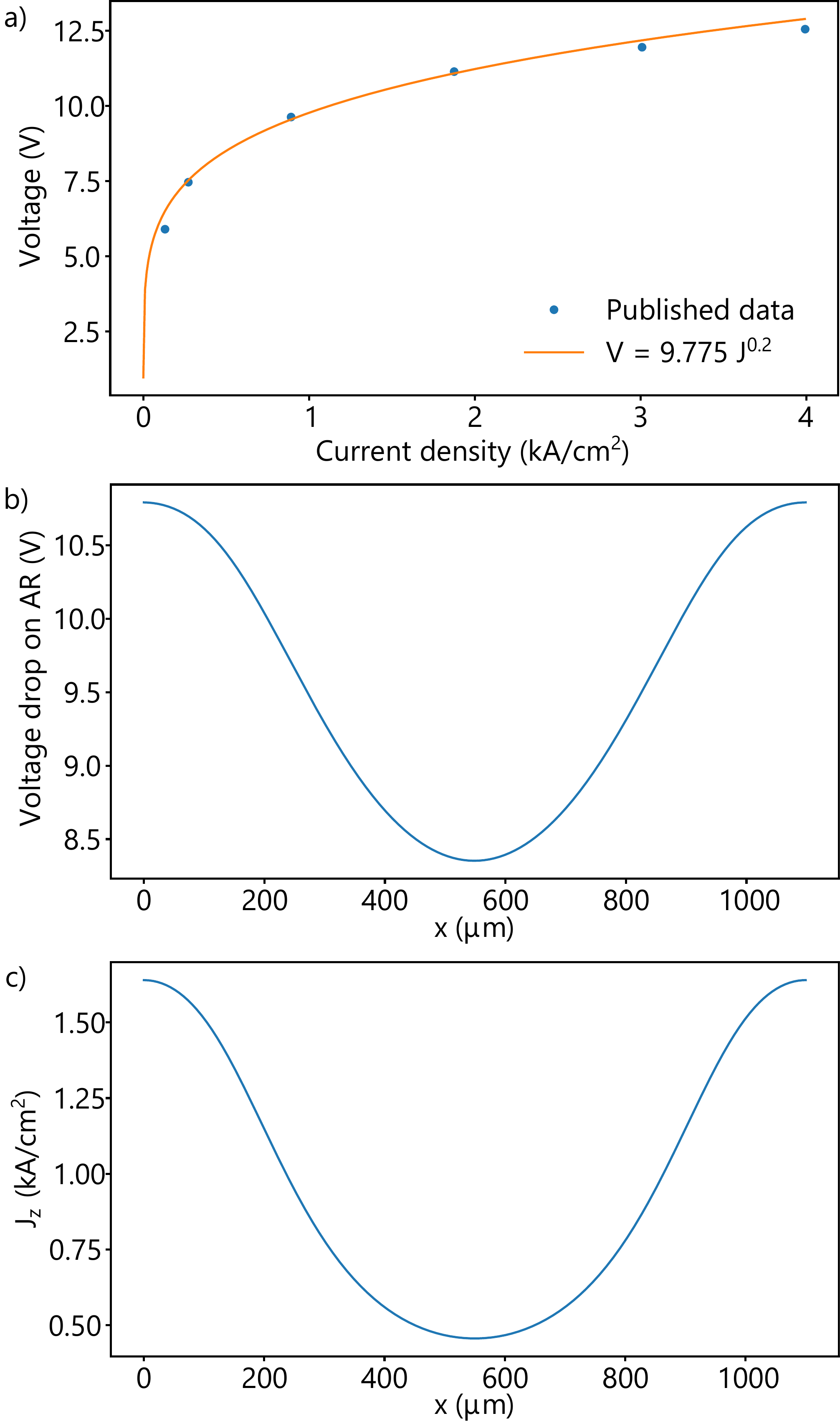}
\caption{(a) V-J fitting of the active region, based on Ref. \citenum{bismuto2010electrically}. (b) Active region voltage drop along the horizontal $x$ axis, with a bias of 20 V on the device, simulated by COMSOL Multiphysics. (c) Vertical current density $J_z$ (at the center layer of the active region) along the horizontal $x$ axis, with a bias of 20 V on the device, simulated by COMSOL Multiphysics. }
\label{fig:COMSOL_1D}
\end{figure}

As mentioned in the main text, a 2D COMSOL simulation with Electric Currents module is performed to investigate the electrical pumping of the active region. The electric potential and current density distribution in the cross-sectional plane of the PhC-QCL are shown in Figs. \ref{fig:COMSOL_2D} (a) and (b). In the COMSOL model, the device width is 1.1 mm. The substrate is 200 \textmu m thick, with the doping of 1.5 $\times 10 ^ {16}$ cm$^{-3}$. The electron mobilities in InP:Si and active region are 2 $\times 10^3$ cm$^2$/V/s and 1 $\times 10^3$ cm$^2$/V/s, respectively \cite{zheng2000electrical}. On top of the substrate, the gold contacts are 100 \textmu m wide, and 50 \textmu m away from the edge, opening the emission window. The bottom cladding is fully covered by the gold. The electric potential of the top and bottom electrodes are set to 0 and -20 V. The vertical conductivity component $\sigma_z$ of the active region is approximated by analytically fitting the data from Ref. \citenum{bismuto2010electrically} (where the laser shares the same active region as the presented PhC-QCL), as shown in Fig. \ref{fig:COMSOL_1D} (a). Figure \ref{fig:COMSOL_2D} (a) and (b) clearly show that the electrical pumping on active region is non-uniform. The voltage drop across the active region relative to the horizontal axis $x$ is illustrated in Fig. \ref{fig:COMSOL_1D} (b). With a bias of 20 V on the whole device, the largest voltage drop across the active region is 10.8 V (close to the edge), whereas at the center the voltage drop is only 8.35 V. The result of current density is shown in Fig. \ref{fig:COMSOL_1D} (c), where the vertical current density $J_z$ distribution relative to the center layer of active region is depicted. The $J_z$ near the edge is around 1.64 kA/cm$^2$, whereas at the center it is only 0.46 kA/cm$^2$. According to the Ref. \citenum{bismuto2010electrically}, the current density threshold for the active region is 1.50 kA/cm$^2$. This means only the 19\% of the active region in this COMSOL model reaches the threshold. If we promote to a 3D case, the result indicates that almost 66\% of the active region is still below threshold, when the laser is biased by 20 V.

\subsection{Edge emission power}
\label{sec:edge}

\begin{figure}[htbp]
\centering
\includegraphics[width=0.75\linewidth]{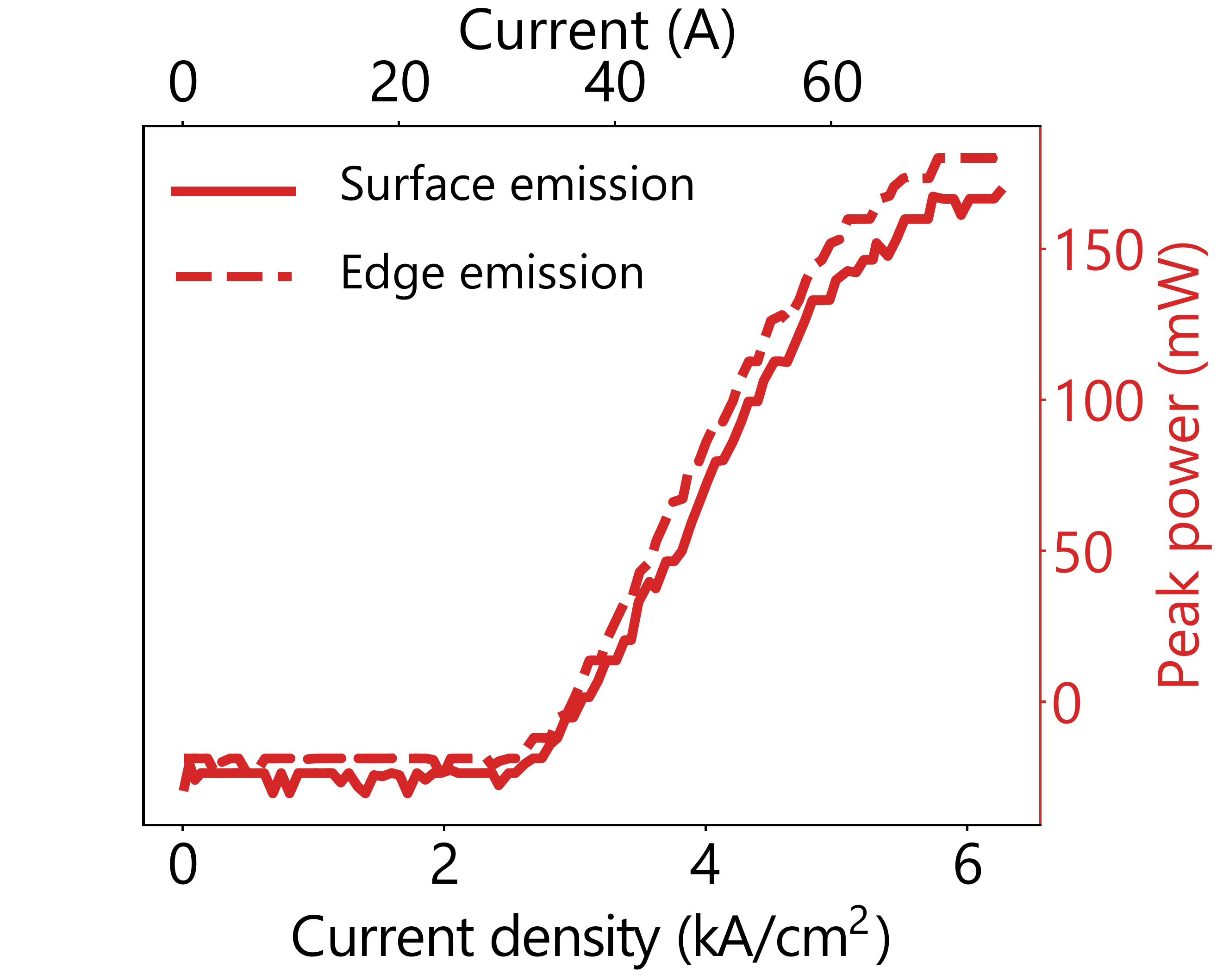}
\caption{Power-current characteristics of a PhC-QCL collected through both edge and surface emission, measured at 288 K under pulsed operation (52 ns at 9.615 kHz). The structural parameters: period $= 2.695$ \textmu m, filling factor $= 0.40$, circular shaped pillars. The power of edge emission is collected from only one of four facets.}
\label{fig:edge}
\end{figure}

Figure \ref{fig:edge} shows the characterization on both the edge and surface emission power of a PhC-QCL at pulsed operation. The laser here is a circular shaped PhC-QCL, with the period $= 2.695$ \textmu m and filling factor $= 0.40$. Due to the symmetry, four edge facets of the laser are supposed to provide similar performances, although only one edge facet can be directly measured. Therefore, the total output power can be almost 5 times the surface emission power, in this case. 

\subsection{Additional data in the band structure measurement}
\label{sec:bandmeas}

\begin{figure}[htbp]
\centering
\includegraphics[width=0.75\linewidth]{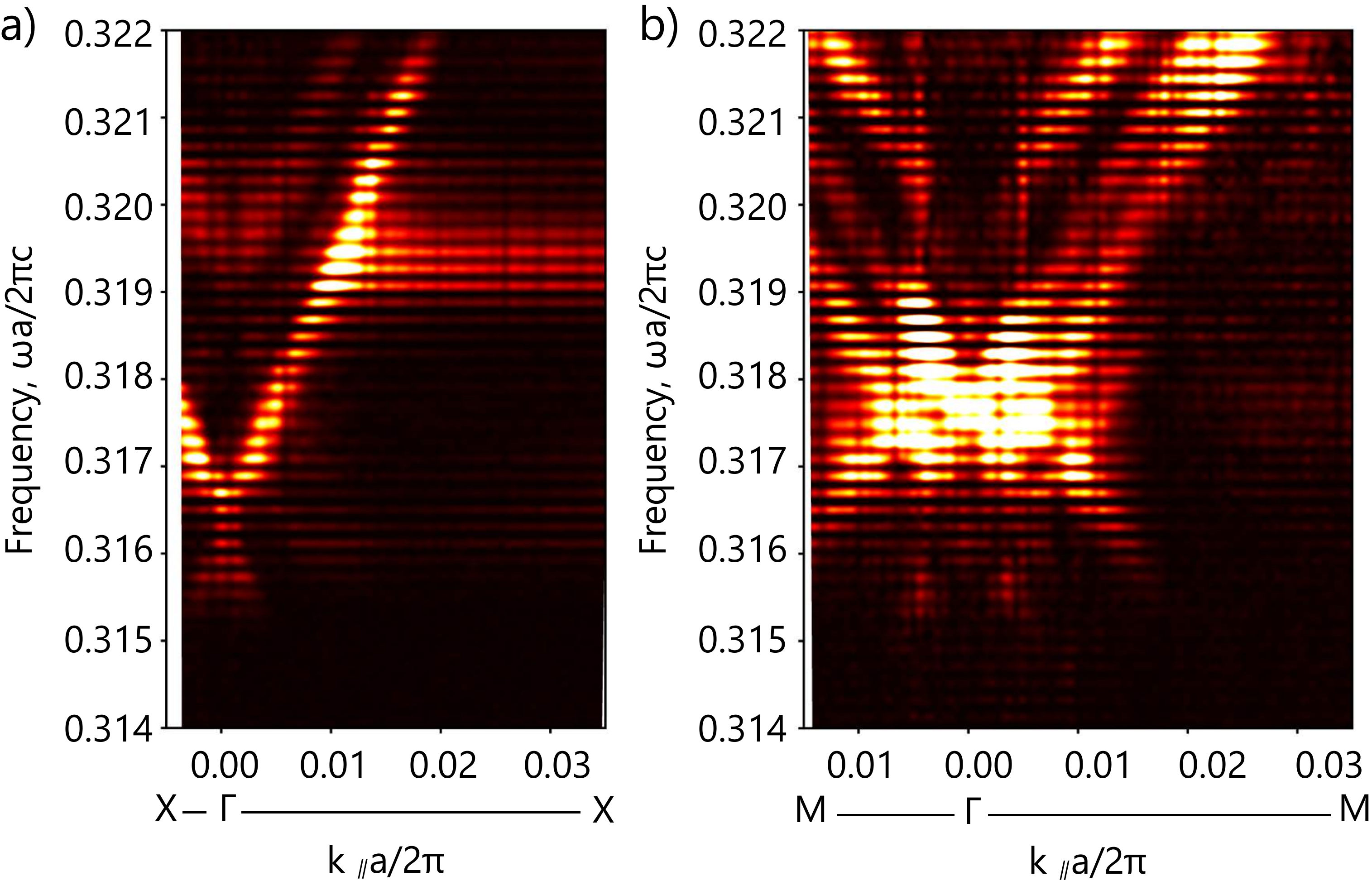}
\caption{Measured bands along $\Gamma$ - X (a) and $\Gamma$ - M (b) directions, respectively (original data of Fig. \ref{fig:4}(c)).}
\label{fig:comp_bands}
\end{figure}

\begin{figure}[htbp]
\centering
\includegraphics[width=0.75\linewidth]{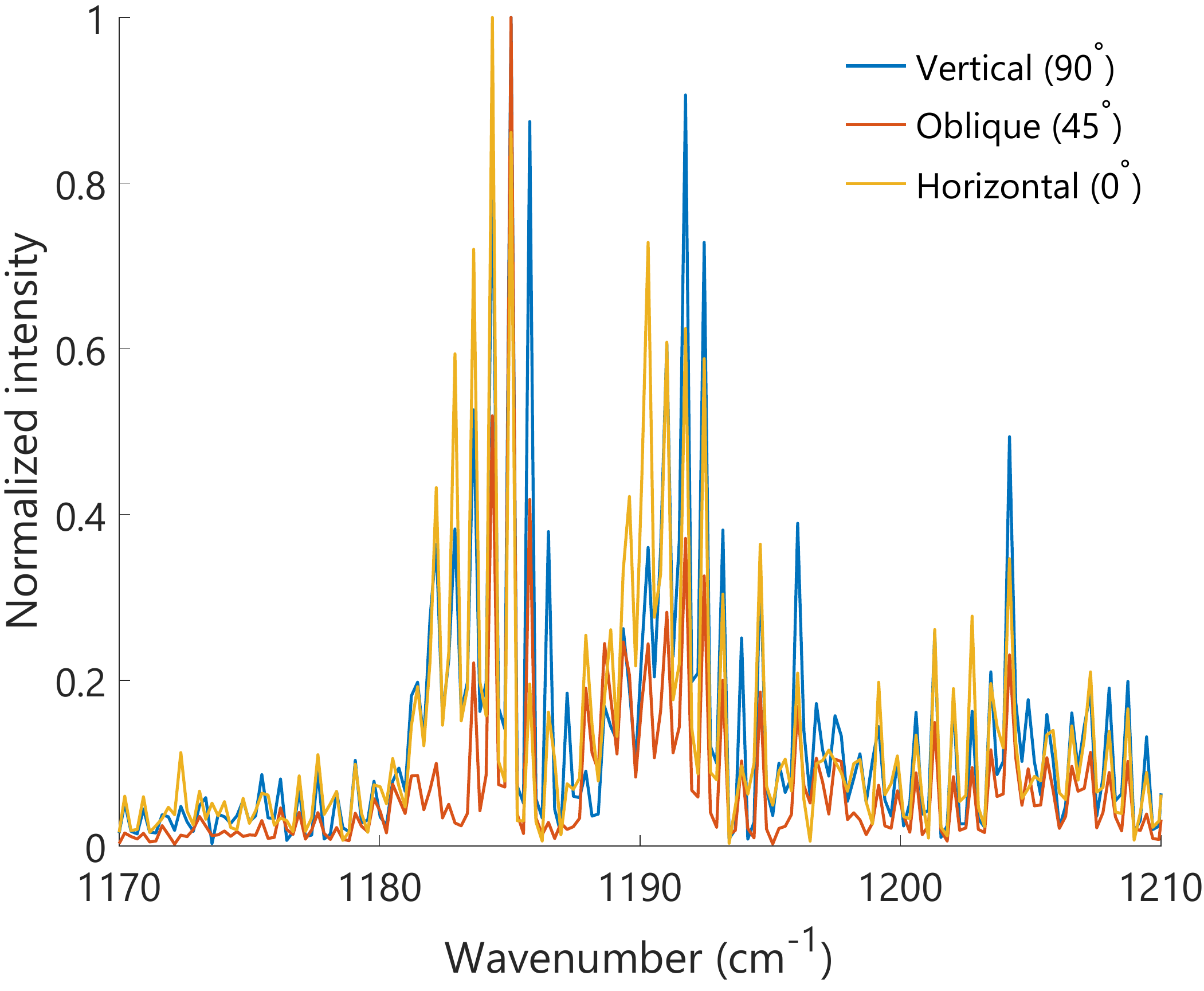}
\caption{Spectra measured using the PhC-QCL as the detector of the FTIR, with the incident light in three polarizations. The PhC-QCL is the same as characterized in Fig. \ref{fig:4}(c) of the main body: circular pillar, period $= 2.680$ \textmu m, filling factor $= 0.40$.}
\label{fig:pol_band}
\end{figure}

Figure \ref{fig:comp_bands} exhibits the complete results of the band measurement of the PhC-QCL along $\Gamma$ - X and $\Gamma$ - M, respectively. The Fig. \ref{fig:4}(c) in the main body is a combination of the results here.

The influence of the incident polarization on the band structure measurement is investigated using a $\lambda$/4 plate and a polarizer in the setup. The $\lambda$/4 plate is placed between the FTIR and the mirror (see Fig. \ref{fig:4}(a) in the main text), changing the incident linearly polarized beam into an elliptically polarized beam (major polarization retains vertical). The polarizer is placed between the mirror and the PhC-QCL detector, selecting the polarization of the incident beam. The PhC-QCL detector is the same one as shown previously in the band structure measurement (the circular pillar PhC-QCL). The PhC-QCL is mounted in the way as $\Gamma$-X scanning (PhC lattices are either vertical or horizontal), but it stays at $\Gamma$ point without any rotation (keeping normal incidence). The temperature of the PhC-QCL is controlled as 291 K. The driving condition of the QCL frequency comb source is the same as shown in the main text (continuous wave operation). Three linear polarizations are selected for comparison: horizontal (0$^{\circ}$), oblique (45$^{\circ}$) and vertical (90$^{\circ}$). The spectral resolution of the FTIR is 0.25 cm$^{-1}$. To compensate for the weaker signal, the integration time of the 45$^{\circ}$ polarization measurement is 4$T$, and the one of the horizontal polarization measurement is 12$T$, where $T$ is the integration time of the vertical polarization measurement. As shown in Fig. \ref{fig:pol_band}, the incident polarization makes a difference on the detected spectrum (especially for the 45$^{\circ}$ measurement). All these spectra are normalized by their individual maximum intensity.


\end{document}